\documentclass[aps,prl,reprint,groupedaddress,longbibliography,nofootinbib]{revtex4-1}

\usepackage{amsfonts}
\usepackage{amsmath}
\usepackage{tikz}
\usetikzlibrary{arrows,calc}
\usetikzlibrary{patterns,snakes}

\newcommand{\bea}{\begin{eqnarray}}
\newcommand{\eea}{\end{eqnarray}}

\begin{document}

%\preprint{}

\title{Mechanism of slow equilibration of isolated quantum systems}


\author{Anatoly Dymarsky}
\affiliation{Department of Physics and Astronomy, \\ University of Kentucky, Lexington, KY 40506}
\affiliation{Skolkovo Institute of Science and Technology, \\ Skolkovo Innovation Center, Moscow, Russia, 143026}
%\thanks{}
%\altaffiliation{Skolkovo Institute of Science and Technology}
%\altaffiliation{University of Kentucky}


\date{\today}
\begin{abstract}

We discuss the approach toward equilibrium of an isolated quantum system. For a wide class of systems we argue that the time-averaged expectation value of a local operator in any initial state is bounded by the so-called deviation function, which characterizes maximal deviation from the equilibrium for all states with a given value of energy fluctuations. We provide numerical evidence that the bound is approximately saturated by the initial configurations with spatial inhomogeneities at macroscopic level. In this way the deviation function establishes an explicit connection between the macroscopically observed  timescales associated with transport and properties of microscopic matrix elements. The form of the deviation function indicates that among the slowest states which saturate the bound there are also those with arbitrarily long equilibration times. %For ergodic systems we conjecture that the integrated variance of the expectation value for any initial state is bounded by Thouless time. 

\end{abstract}

%\keywords{}

\maketitle
%One of the central open questions is that of dynamics of equilibration.

Emergence of statistical mechanics for isolated quantum systems has been an active topic of research for the last decade.  While thermodynamic equilibrium of a quantum many-body system can be successfully explained using typicality arguments \cite{goldstein2006canonical,popescu2006entanglement,reimann2007typicality}, to quantitatively describe the approach towards equilibrium proved to be a far more challenging task. Eventual equilibration of quantum systems was proved under a wide range of conditions in \cite{reimann2008foundation,linden2009quantum,goldstein2010long,linden2010speed,short2011equilibration,reimann2012equilibration,reimann2015generalization} (also see \cite{gogolin2016equilibration,wilming2018equilibration} for reviews), but the resulting uniform bound on equilibration timescale established so far is exponential in system size \cite{short2012quantum}. Clearly, this is not the timescales observed macroscopically. At the same time much stronger bounds independent of system size were found for the special initial conditions or observables, either typical in a technical sense or with low entanglement, etc.~\cite{malabarba2014quantum,goldstein3380approach,goldstein2015extremely,reimann2016typical,balz2017typical,de2018equilibration,cramer2012thermalization,garcia2017equilibration}. 
These results can explain local pre-thermalization of quantum system, but fall short of describing late time dynamics normally associated with transport. In the latter case characteristic timescales normally grow polynomially with the system size. In this paper we elucidate the mechanism of slow equilibration, when characteristic times grow with the system size, and identify relevant timescales microscopically, starting from the underlying quantum-mechanical description.  

%{\it Setup and notations.}
In this work we consider equilibration from the point of view of an observable. It is known that there are always artificial observables which require exponentially long in system size time to thermalize \cite{goldstein2013time,malabarba2014quantum}. Our focus instead is on the physically motivated observables, such as macroscopic averaged or local observables which are expected to equilibrate at most in polynomial time. % In what follows we do not restrict the observable $A$ to be of any particular type. %, with the only requirement for it to be bounded. 
We  assume that the quantum system under consideration has a discrete spectrum of energies $E_i$ and corresponding eigenstates $|E_i\rangle$. As a technical assumption which can be relaxed we assume the spectrum is non-degenerate. 
The initial states $\Psi$ will belong to a microcanonical interval specified by a mean energy $E_\Psi$ and width $2\Delta E_\Psi$, 
\bea
\label{micro}
|\Psi\rangle =\sum_i c_i |E_i\rangle, \quad E_i \in [E_\Psi-\Delta E_\Psi,E_\Psi+\Delta E_\Psi].
\eea 
For the given initial $\Psi$ equilibrated value of $A$ is simply 
\bea
\overline{\langle \Psi|A|\Psi\rangle}=\lim_{T\rightarrow \infty} {1\over T}\int_{0}^T 
{\langle \Psi|A(t)|\Psi\rangle}dt=\sum_i |c_i|^2 A_{ii}. \ \ 
\eea
The deviation of $A$ from equilibrium is captured by an auxiliary operator $\hat{A}$, that has the following form in the energy eigenbasis
\bea
\hat{A}_{ij}=\left\{ \begin{array}{cc}
A_{ij}, & i\neq j \\
0, & i=j
\end{array}\right. .
\eea
We assume $A$ is bounded, $||{A}||=1$, thus without loss of generality $\hat{A}$ is bounded as well.
Finally, we introduce time-dependent deviation from the equilibrium 
\bea
\nonumber
a(t)&:=&\langle \Psi|\hat{A}(t)|\Psi\rangle=\langle \Psi|{A}(t)|\Psi\rangle-\overline{\langle \Psi|A|\Psi\rangle}=\\
&&\sum_{i\neq j} c_i^* c_j^{} A_{ij} e^{-i(E_i-E_j)t}. \label{at}
\eea

The approach of $a(t)$ to zero can be explained from \eqref{at} through dephasing. The crucial question is to identify characteristic energy difference $\delta E\sim |E_i-E_j|$ which governs time dynamics. Clearly, $\delta E$ can not be larger than the energy variance of the initial state, but the latter is normally growing with the system size, which seemingly should result in vanishing equilibration times. It was suggested in \cite{de2018equilibration} that $\delta E$ is bounded by the form of the matrix elements $A_{ij}$, which for local $A$ approach zero exponentially quickly for large  $|E_i-E_j|\sim 1$. This can explain why characteristic equilibration timescales do not decrease for large system sizes, but fail to explain why they actually increase.  To elucidate the approach of $a(t)$ to zero, we first consider a qualitative argument. We start with an initial out of equilibrium state \eqref{micro} with large $\Delta E_\Psi$ and $a(0)\neq 0$ and consider $a(t)$ at some late time $t$.  The initial state  can be represented as a sum of $\Delta E_\Psi\, t/(2\pi)$ states of the form \eqref{micro}, each confined to its own interval of size $2\pi/t$ centered at $E_k=E_\Psi+(2k-1)\pi/t+\Delta E_\Psi$, $\Psi=\sum_{k=1}^{\Delta E_\Psi t/2\pi}\Psi_k$.
Then qualitatively we can assume that different $\Psi_k$ are mutually dephased, such that 
%Then for all pairs $E_i,E_j$ such that $|E_i-E_j|t\gtrsim 1$ we can assume that the corresponding terms in \eqref{at} have random phase and average to zero.
\bea
\label{sumk}
a(t)=\langle \Psi|\hat{A}(t)|\Psi\rangle \approx \sum_{k=1}^{\Delta E_\Psi t/2\pi} \langle \Psi_k|\hat{A}(t)|\Psi_k\rangle.
\eea 
Each term in  \eqref{sumk} is bounded in terms of the deviation functions $\hat{x}_{\rm max/min}(E,\Delta E)$ which are defined as the maximal/minimal eigenvalues of matrix $\hat{A}_{ij}$ projected on the interval $[E-\Delta E,E+\Delta E]$. These functions are closely related to the deviation functions ${x}_{\rm max/min}(E,\Delta E)$ introduced in \cite{dymarsky2017canonical} with the only difference being that in the latter case the corresponding matrix is $A_{ij}$. (In other words $\hat{x}$ and $x$ differ only by the inclusion of the diagonal elements $A_{ii}$ and in the quantum ergodic case functions $\hat{x}$ and $x$ are the same up to unimportant volume-suppressed corrections. In the integrable case these functions are different. A more detailed comparison of $x$ and $\hat{x}$ can be found in the Appendix.)
Assuming functions $\hat{x}$ smoothly depend on the mean energy of the interval, all $E_k$ can be substituted by  $E_\Psi$ and we find 
\bea
\label{naivebound}
\hat{x}_{\rm max}(E_\Psi,\pi/t)\gtrsim a(t)\gtrsim \hat{x}_{\rm min}(E_\Psi,\pi/t).
\eea

\begin{figure}
\begin{tikzpicture}[scale=1]
 \draw[lightgray,fill] (1,1)  -- (3.5,1) -- (1,3.5);
 \draw[lightgray,fill] (6,6)  -- (3.5,6) -- (6,3.5);
 \draw[] (1,1) -- (1,6) --  (6,6) -- (6,1) -- (1,1);
% \draw[lightgray,fill] (1,3.5)  -- (3.5,1) -- (6,1) --(6,3.5)--(3.5,6)--(1,6);
 \draw[] (3.5,1) -- (3.5,6);
 \draw[] (1,3.5) -- (6,3.5);
 \node[] at(3.5,0.6) {$E$};
 \node[] at(6,0.6) {$E+\Delta E$};
 \node[] at(1,0.6) {$E-\Delta E$};

\end{tikzpicture}
\caption{Schematic visualization of $\hat{A}_{ij}$ projected on the energy interval $[E-\Delta E, E+\Delta E]$. 
Functions $\hat{x}_{\rm max}$ and $\hat{x}_{\rm min}$ are  defined as the maximal/minimal eigenvalues of this matrix. Functions $\tilde{x}_{\rm max}$ and $\tilde{x}_{\rm min}$ are defined as the maximal/minimal eigenvalues of the same matrix after all matrix elements in the gray area substituted by zero, see eq.~\eqref{bandana}.}
\label{fig:difference}
\end{figure}
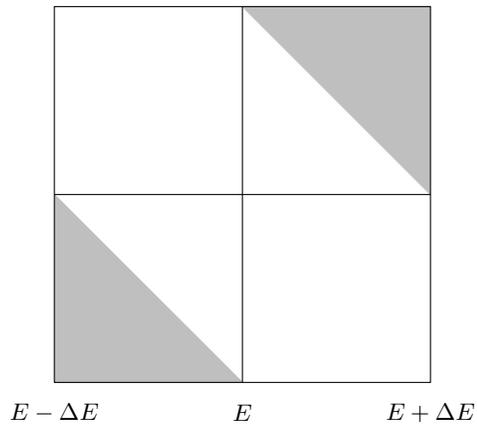

This bound can not uniformly apply to all states even at late times. We show in the Appendix  that it is always possible to find an initial state such that different $\Psi_k$ become mutually coherent at any given point in time $t={t^*}$ and both $a(0)$ and $a({t^*})$ are far from being zero.  Nevertheless a similar bound applies to the time-averaged $a(t)$,
\bea
\label{defat}
\langle a\rangle_T=\int_{-\infty}^{\infty} dt\, {\sin(\pi t/T)\over \pi t} a(t).
\eea
The time average $\langle a\rangle_T$ can be bounded by the extreme eigenvalues $\tilde{x}_{\rm max/min}(E,\Delta E)$ of the band matrix $\hat{A}_{\Delta E}$, 
\bea
\label{bandana}
(\hat{A}_{\Delta E})_{ij}=\left\{ 
\begin{array}{r c}
\hat{A}_{ij}, & |E_i-E_j|\leq \Delta E, \\
0, &|E_i-E_j|\geq \Delta E,
\end{array}\right. \\
E_i,E_j\in[E-\Delta E,E+\Delta E]. \label{interval}
\eea
To clarify, $\hat{x}$ stands for the extreme eigenvalues of matrix $\hat{A}_{ij}$ restricted to the interval \eqref{interval}, while $\tilde{x}$ stands for  extreme eigenvalues of the same matrix after taking certain off-diagonal elements to be zero. 
The difference in definitions of $\hat{x}$ and $\tilde{x}$ is visualized in Fig.~\ref{fig:difference}. It was shown in \cite{dymarsky2018bound} in full generality that for any initial $\Psi$ and  $T$ 
\bea\label{rigorous}
2\tilde{x}_{\rm max}(E',{\pi/T})+\hat{x}_{\rm max}(E'',\pi/(2T)) \geq \langle a\rangle_{T}, 
\eea
and a similar inequality bounds $\langle a\rangle_{T}$ from below. Energies $E'$ and $E''$ have to belong to the original interval \eqref{micro} associated with $\Psi$. Assuming that $\hat{x}$ and $\tilde{x}$ smoothly depend $E$ only through the effective temperature, so far the original state has relatively small energy variance $\Delta E_\Psi\ll E_\Psi$ one can substitute $E',E''$ by $E_\Psi$.

A priory $\hat{x}$ and $\tilde{x}$ are not directly related. The former quantity measures maximal deviation from the equilibrium $a(0)$ for all states of the form \eqref{micro} with $E_\Psi=E$ and $\Delta E_\Psi=\Delta E$, while the latter measures maximum value of $\langle a \rangle_{\pi/\Delta E}$ for the same class of initial states. In case $a(t)$ is oscillating exactly with the period $\pi/\Delta E$, $|\tilde{x}|$ can be much larger than $|\hat{x}|$. At the same time for a wide class of systems we expect $a(t)$ not to exhibit persistent oscillations with the time periods equal to or exceeding the timescales associated with the macroscopic transport. Thus, for quantum ergodic systems, for which all off-diagonal matrix elements of local observables were verified to be exponentially small \cite{dymarsky2016subsystem}, we expect that substituting certain matrix elements with zero can not parametrically increase extreme eigenvalues. Hence for sufficiently small $\Delta E$, we expect $\tilde{x}$ can be bounded in terms of $\hat{x}$,
\bea
\label{inequality1}
\tilde{x}_{\rm max}(E,\Delta E)&\leq &\kappa\, \hat{x}_{\rm max}(E,\Delta E),\\
\tilde{x}_{\rm min}(E,\Delta E)&\geq &\kappa\, \hat{x}_{\rm min}(E,\Delta E).
\label{inequality2}
\eea 
Here $\kappa$ is some constant, which is $\Delta E$ and system size-independent. This bound with different values of $\kappa$ may apply more generally e.g.~to integrable systems, as we demonstrate numerically below.   In all cases considered $\kappa\leq 1$. 
Using monotonicity of $\hat{x}$ as a function of $\Delta E$, we find that for sufficiently large $T$ (compare with \eqref{naivebound}),
\bea
\label{bound}
\kappa' \hat{x}_{\rm max}(E_\Psi,\pi/T) \geq \langle a\rangle_T \geq \kappa' \hat{x}_{\rm min}(E_\Psi,\pi/T),
\eea 
where $\kappa'\leq 3 \kappa$ is some constant of order $1$. This bound applies to all initial states with $\Delta E_\Psi \ll E_\Psi$ and for times $T$ not exceeding Heisenberg time. 

\begin{figure}
\includegraphics[width=0.45\textwidth]{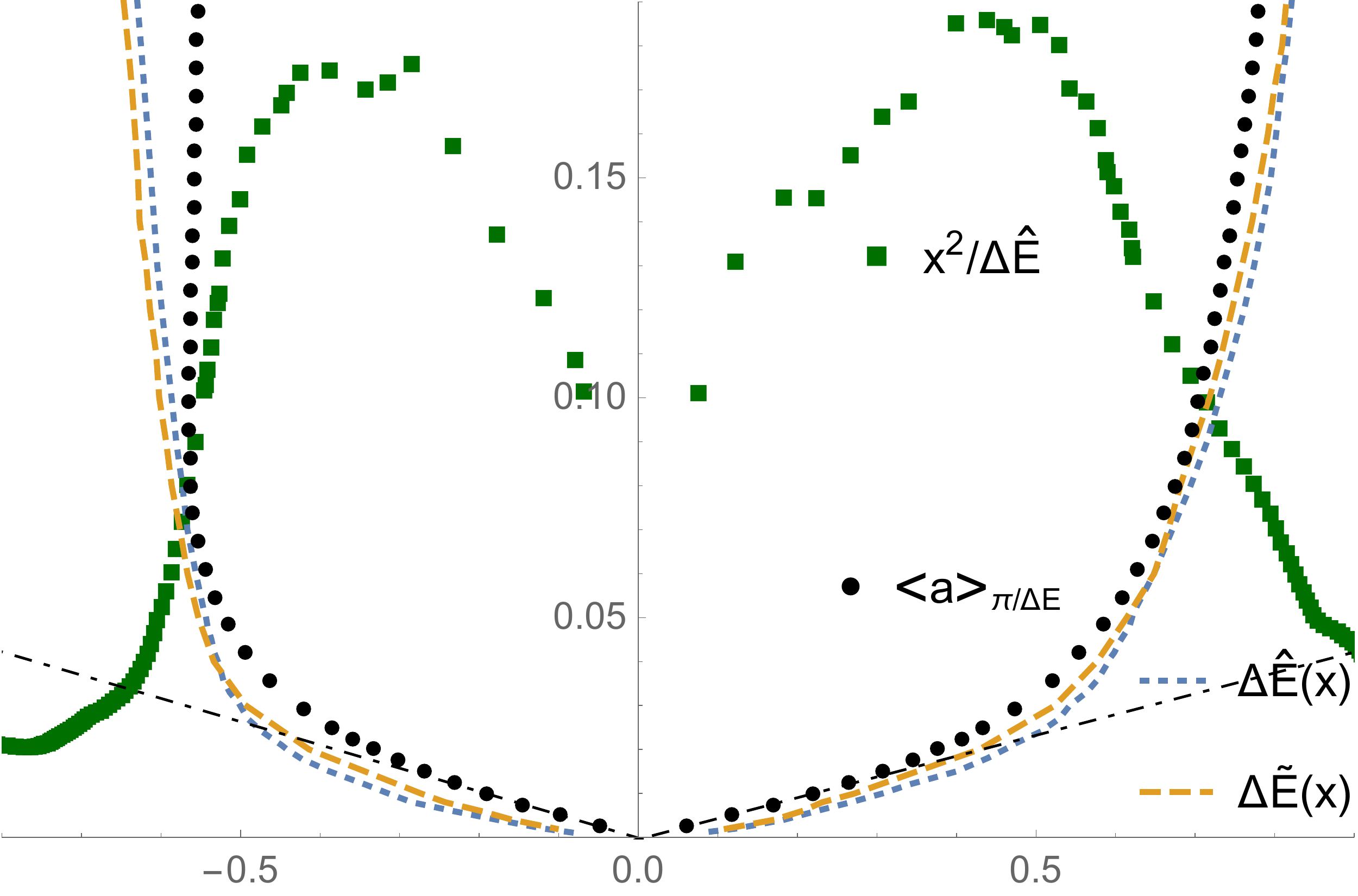}
\caption{Plots of $\Delta \hat{E}(E,x)$ (blue dotted line) and $\Delta \tilde{E}(E,x)$ (yellow dashed line) for $A=\sigma_x^1$, $h=0.7$, $L=17$, $E=0$, superimposed with the plots of $x=\langle a\rangle_T$ as a function of $\Delta E=\pi/T$ (black lines) for $\ell=9, e_1=-11.44, e_2=10.63$ (left) and $\ell=9, e_1=, 12.02, e_2=-14.46$ (right). Straight black dotted lines is the linear fit of $\langle a\rangle_{\pi/\Delta E}$ for small $\Delta E$. Their values at $x=\pm 1$ give the approximate value of $\pi/(2\tau_D)\approx 0.04$.
Green squares in the background show $(5/L^2){x}^2/\Delta \hat{E}$ as a function of $\hat{x}$. The overall scaling coefficient $5/L^2$ is chosen by convenience to fit  $x^2/\Delta \hat{E}$ on the same plot with $\Delta \hat{E}(x)$.} 
\label{fig:E07}
\end{figure}

Below we provide numerical evidence for (\ref{inequality1},\ref{inequality2}) and \eqref{bound}  in case of an open Ising spin-chain in one dimensions,
\bea
\label{H}
H=-\sum_{k=1}^{L-1} \sigma_z^k \otimes \sigma_z^{k+1}+g\sum_{k=1}^L \sigma_x^k+h\sum_{k=1}^L\sigma_z^k.
\eea
We take $g=1.05$ and explore different regimes: integrable  $h=0$ and quantum ergodic $h=0.4,h=0.7$. In the latter case the system is known to exhibit diffusive transport \cite{kim2013ballistic}. We consider several different operators, local ones $A=\sigma_x^1$ and $A=\sigma_z^1$, and averaged quantities  $A=\sum_{k=1}^L \sigma_x^k/L$, $A=\sum_{k=1}^L \sigma_z^k/L$.  
In Fig.~\ref{fig:E07} we show functions $\hat{x}$ and $\tilde{x}$ for $A=\sigma_x^1$ and $E=0,h=0.7$. For convenience two functions $\hat{x}_{\rm \min/\rm max}(E,\Delta E)$ are represented in terms of the inverse function $\Delta \hat{E}(x)$ (from now on we do not write $E$ explicitly), 
\bea
\Delta \hat{E}(\hat{x}_{\rm max/ min}(\Delta E))=\Delta E,
\eea
and similarly $\Delta \tilde{E}(x)$ for $\tilde{x}$. Numerics shows, that the function $\Delta \tilde{E}(x)$ is approximately equal to $\Delta \hat{E}(x)$ up to $\Delta E\lesssim 0.4$. Importantly, $\Delta \tilde{E}(x)\geq \Delta \hat{E}(x)$ up to $\Delta E\lesssim 0.04$. This is already enough to guarantee the inequality \eqref{bound} for $T\ge \tau$, where ${\pi/ \tau}=0.04$ and $\kappa'=3$. In reality we may expect the inequality to be satisfied already for $T\ge \pi/0.4$ and smaller $\kappa'$.

An important observation is that $\tau$ can be associated with the diffusion (Thouless) time. To make this connection we consider time-evolution of different configurations with macroscopic gradient of energy density. Namely, we represent \eqref{H} as a sum of two interacting spin-chains $H=H_L-\sigma_z^{\ell}\otimes \sigma_z^{\ell+1}+H_R$ and
consider the ``square wave'' configurations $\Psi=|e_1\rangle \otimes |e_2\rangle$, where $|e_1\rangle$ and $|e_2\rangle$ are the eigenvalues of $H_L, H_R$ correspondingly. An analogous setup was considered in \cite{varma2017energy}. We consider $\ell=8,9$ and choose energies $e_1,e_2$ such that the initial deviation $a(0)$ for $A=\sigma_x^1$ is large and the total energy $E_\Psi$ is close to zero (this is done to ensure $\Psi$ belongs to the middle of the spectrum to minimize finite-size effects). 
We plot $\langle a \rangle_T$ as a function of $T=\pi/\Delta E$ for several selected $\Psi$ in Fig.~\ref{fig:E07}.
For large $T$ exceeding Thouless time the integral \eqref{defat} can be approximated as  (here we also assume $a(t)$ is an even function of $t$)
\bea
\label{thermtime}
\langle a\rangle_T\approx 2 \int_0^\infty a(t) dt/T. 
\eea
Using late time approximation $a(t)\approx a(0)e^{-t/\tau_D}$ we find $\langle a\rangle_T \approx 2\tau_D/T$, where $\tau_D$ is the diffusive, or Thouless, time. 
Hence the slope of $\langle a\rangle_{T}$,   as a function of $\Delta E=\pi/T$ can be associated with $\tau_D$. The plots in Fig.~\ref{fig:E07} demonstrate that $2\tau_D$ is approximately equal to $\tau$. 
Identifying $\tau$ as Thouless time provides the following interpretation: the inequalities \eqref{bound} uniformly apply to all states for times $T$ equal to or exceeding timescales of macroscopic equilibration. 

Remarkably the inequalities \eqref{bound}  with $\kappa'=1$ are approximately saturated at $T\sim \tau/2$ by the square-wave configurations with largest amplitudes  (see Fig.~\ref{fig:E07}). The points of saturation approximately coincide with the points of largest curvature of  $\Delta \hat{E}(x)$, $x_+$ and $x_-$.  It can be also shown analytically that $\Delta \hat{E}(x=\pm 1)$ remains finite in the limit $L\rightarrow \infty$.
Combining this with the monotonicity of $\hat{x}_{\rm max/min}$ as a function of $\Delta E$ suggests  the following behavior of $\Delta \hat{E}(x)$  in the thermodynamic limit.
As $L$ increases $\Delta \hat{E}(x)$ will remain finite for $-1\le x< x_{-}$ and $x_{+}< x\le 1$ and will go to zero at least as $1/L^2$ for $x_{+} \ge x \ge x_{-}$. Hence Thouless energy can be {\it defined} as the value of $\Delta \hat{E}(x)$ at the boundary points of the region $[x_-,x_+]$, within which $\Delta \hat{E}(x)$ approaches zero in the thermodynamic limit. 

It is instructive to discuss the behavior of  $\Delta \tilde{E}(x)$ and $\Delta \hat{E}(x)$ at small $x$. It can be shown in full generality that for $\Delta E$ much smaller than the inverse  macroscopic equilibration times
\bea
\label{parabola}
x^2\ge  {\Delta Ef^2(0)\over 4\pi}\ ,\\
f^2(0):= \lim_{\omega \rightarrow 0} \int_{-\infty}^\infty dt  e^{-i \omega t} \langle A(t)A(0)\rangle. \label{fdef0}
\eea
The autocorrelation function in \eqref{fdef0} is calculated over the microcanonical window $[E-\Delta E/2, E+\Delta E/2]$ (see Appendix for details).
For most observables $f^2(0)$ remains finite or grows with the system size in the thermodynamic limit (for a diffusive quantity in 1D $f^2(0)\sim L$ \cite{d2016quantum,luitz2016anomalous}), although there are observables, e.g. the operators of the form $A=i[H,B]$, for which $f^2(0)=0$. Such observables are excluded from further consideration. 
The inequality \eqref{parabola} means that near the origin $\Delta \hat{E}(x)$, $\Delta \tilde{E}(x)$ grow slowly, not faster than a parabola. 

The bound \eqref{bound} is approximately saturated at $T=T^*$ by the state \eqref{micro} which maximizes the deviation $\langle \Psi|\hat{A}|\Psi\rangle$ for the given value of $\Delta E_\Psi=\pi/T^*$. We would call such states the ``slowest.'' For such states $a(0)=\hat{x}(\Delta E_\Psi)$ and $|\dot{a}(t)|\leq 2|a(0)|\Delta E_\Psi$, which gives $\langle a\rangle_{T^*} \gtrsim \hat{x}(\pi/T^*)$. If we define equilibration time for such states as the value of $T\langle a\rangle_T$ for very large $T$,  it will be at least of the order $|a(0)|/\Delta E_\Psi\gtrsim f(0)/\Delta E_\Psi^{1/2}$. By choosing $\Delta E_\Psi$ sufficiently small one can make equilibration time arbitrarily long. Thus, in full generality a diffusive system admits a family of  states which remain out of equilibrium parametrically longer than Thouless time. It remains to be seen what role these states may play at the macroscopic level.

Finally, we discuss another characteristic of equilibration dynamics, the integrated expectation value variance 
\bea
\label{newt}
\mathcal{T}=\int_0^{\tau_\infty} a(t)^2 dt.
\eea
%Here $\tau_\infty$ is chosen to be much  larger than all macroscopic timescales, but is much smaller than the Heisenberg time. $\mathcal T$ can be interpreted as equilibraion time. It was shown in \cite{short2012quantum,reimann2012equilibration}, that $\mathcal{T}$ is uniformly bounded for a wide class of initial states, but the bound is exponentially large in system size. 
%%While eventual approach of $\langle a \rangle_T$ to zero signals weak equilibration \cite{banuls2011strong}, finite value of $\mathcal{T}$ indicates equilibration of the observable in the conventional sense. 
%%It was shown in great generality in \cite{short2012quantum,reimann2012equilibration}, that $\mathcal{T}$ is finite for all states with a sufficiently large inverse participation ratio, a condition, all states discussed previously would satisfy when the system is quantum ergodic. The weakness of this result is that
%%the uniform bound on $\mathcal{T}$ is exponentially large in system size.   
%We conjecture that for quantum ergodic systems and a diffusive $A$ a tight bound is much smaller and is given by Thouless time, i.e.~it is polynomial in system size. Our argument is based on two assumptions. 
We estimate $\mathcal{T}$ for the ``slowest'' states introduced above as the initial amplitude-squared $|a(0)|^2$ multiplied by the characteristic time $|a(0)|/|\dot{a}|\sim T^*=\pi/\Delta E_\Psi$ necessary for $a(t)$ to approach zero. This gives 
\bea
\label{bigT}
\mathcal{T} \geq \mathcal{T}_0(\Delta E_\Psi),\quad \mathcal{T}_0:={\hat{x}^2(\Delta E) \over 6\Delta E}.
\eea
%Our second assumption is that the inequality \eqref{bigT} is close to be saturated, i.e.~the ``slowest'' states equilibrate  within the characteristic timescale introduced by $\Delta E$.  
In the ergodic case matrix $\hat{A}_{ij}$ inside a sufficiently small energy interval is expected to be random \cite{dymarsky2018bound} causing \eqref{parabola} to be saturated up to a numerical factor. Hence, for small $\Delta E$, $\mathcal{T}_0 \approx f^2(0)\sim L$. When $\Delta E$  becomes larger, of order of Thouless energy, $\hat{x}\sim x_\pm\sim 1$ and   $\mathcal{T}_0 \approx \tau_D\sim L^2$. This is the largest value of $\mathcal{T}_0$. For larger $\Delta E$,  $\mathcal{T}_0$ decreases and eventually goes to zero. This behavior, with two clear peaks at $x\sim x_\pm$ is shown in Fig.~\ref{fig:E07}, where we plot $x^2/\Delta \hat{E}(x)$ as a function of $x$. To summarize, for an ergodic quantum many-body system Thouless time can be defined as the maximal value of $x^2/\Delta E(x)$. It is also interesting to see if Thouless time can impose a universal upper bound on $\mathcal T$ \eqref{newt} for all initial states provided the ergodic system exhibits only one type of transport. 

\begin{figure}
\includegraphics[width=0.45\textwidth]{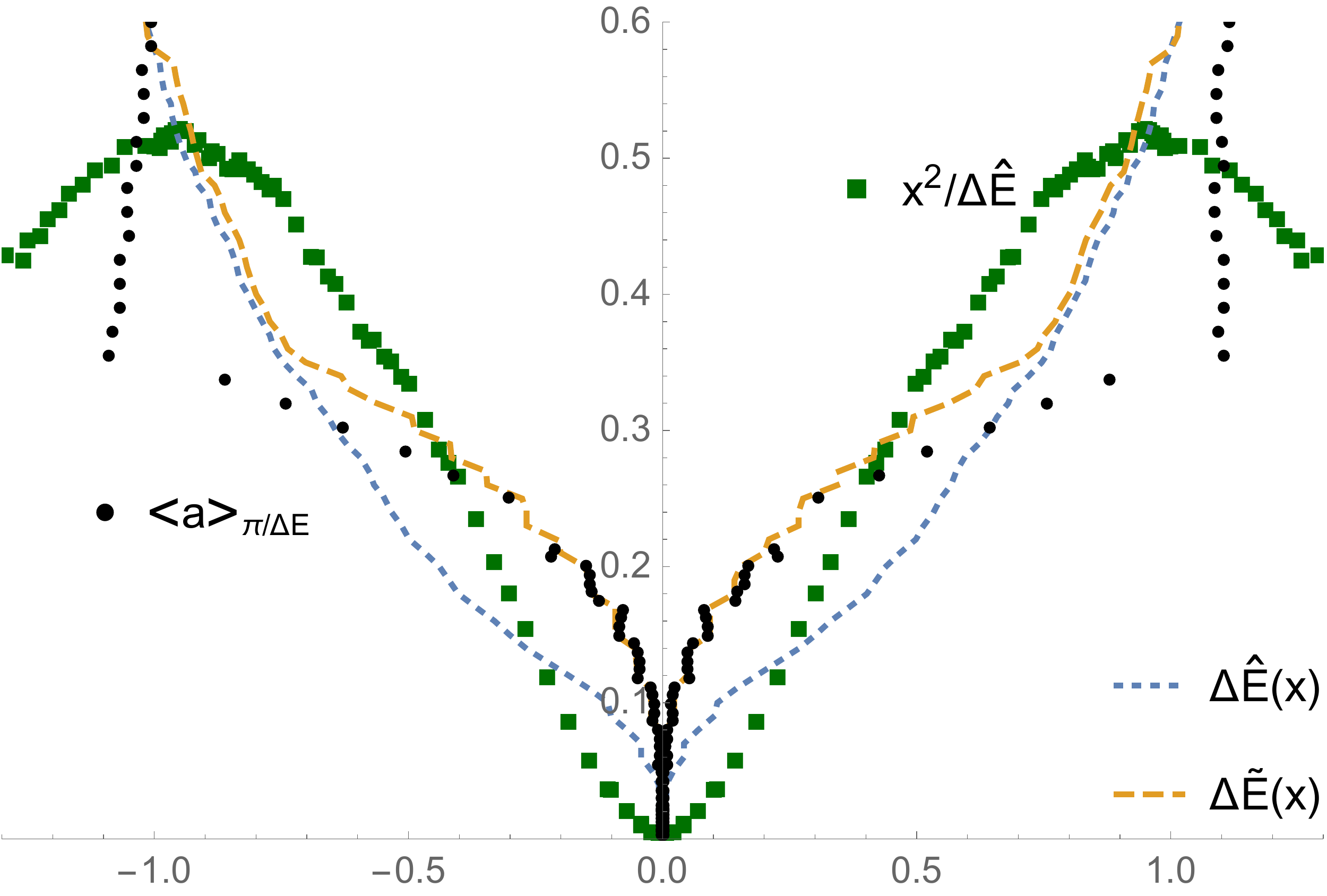}
\caption{Plots of $\Delta \hat{E}(E,x)$ (blue dotted line) and $\Delta \tilde{E}(E,x)$ (yellow dashed line) for $A=\sigma_x^1$, $h=0$, $L=17$, $E=0$, superimposed with the plots of $x=\langle a\rangle_T$ as a function of $\Delta E=2\pi/T$ (black lines) for $\ell=9, e_1=-11.45, e_2=10.14$ (left) and $\ell=8, e_1=10.14, e_2=-11.45$ (right). 
Green squares in the background show $(5/L){x}^2/\Delta \hat{E}$ as a function of $\hat{x}$. The overall scaling coefficient $5/L$ is chosen by convenience to fit  $x^2/\Delta \hat{E}$ on the same plot with $\Delta \hat{E}(x)$.} 
\label{fig:E0}
\end{figure}

Next, we discuss the integrable case $h=0$. The corresponding plots for $A=\sigma_x^1$ are shown in Fig.~\ref{fig:E0}. First we notice that $\Delta \hat{E}(x) \leq \Delta \tilde{E}(x)$ for  $\Delta E\leq \pi/\tau\approx 0.6$. The plots of $\langle a\rangle_T$ as a function of $T$ for macroscopic configurations reach their maximal values at $T\sim 2\tau$. This time can be identified as the ballistic time. Hence the inequality \eqref{bound} is satisfied already for times  somewhat smaller than the macroscopic equilibration timescales. The plot of $x^2/\Delta \hat{E}$ exhibits two peaks, as in the ergodic case, but in the integrable case we expect the value of the peaks to scale as $L$, consistent with the ballistic transport. 

We note that $\langle a\rangle_T$ as a function of $\Delta E=\pi/T$ becoming larger than $\tilde{x}(
\Delta E)$ and $\hat{x}(\Delta E)$ is not a violation of \eqref{bound}, because of  $\kappa'$. It is interesting that for small $\Delta E$, the behavior of $\langle a\rangle_T$ essentially coincides with $\tilde{x}(\Delta E)$. This could mean that for sufficiently late time \eqref{bound} is satisfied with $\kappa'=1$.

The behavior of $\Delta \hat{E}(x)$ at the origin implies that $f^2(0)$ for  $A=\sigma_x^1$ vanishes. The same behavior is exhibited by some other, but not all, local operators. Following \cite{spohn2017interacting}, we attribute this to the corresponding integrable system being non-interacting. 

The  operators  $A=\sigma_z^1, A=\sum_{k=1}^L \sigma_x^k/L$, and $A=\sum_{k=1}^L \sigma_z^k/L$ are discussed in the Appendix. In all cases considered the inequalities (\ref{inequality1},\ref{inequality2}) are satisfied for $\Delta E$ equal or larger than those in the case of $A=\sigma_x^1$, ensuring \eqref{bound} is satisfied.

{\it Discussion.}
In this work we outlined a microscopic mechanism of ``slow'' equilibration when characteristic timescales grow with the system size. 
Qualitatively the expectation value in a state $\Psi$ averaged over time T is given by the expectation value in the state $\Psi$ projected on an energy interval of size $2\pi/T$.  Then the longest macroscopic timescale is determined by the size of the smallest energy interval necessary to achieve a large deviation from the equilibrium. This picture is made precise by the bound \eqref{bound}, which constraints time-averaged expectation value in terms of the deviation functions $\hat{x}_{\rm max/min}(\Delta E)$. %The deviation function is a property of an observable and specifies the extremal deviations from the equilibrium for a given value of energy fluctuations $\Delta E=\Delta E_\Psi$. 
%The bond \eqref{bound} is approximately saturated by the initial configurations with spatial inhomogeneities at macroscopic level,
We identified the longest macroscopic equilibration timescale (Thouless time in the diffusive case) as the maximal value of ${x}^2/\Delta \hat{E}(x)$, or, equivalently, the inverse value of $\Delta \hat{E}(x)$ at the boundary points $x_\pm$ of the interval, within which $\Delta \hat{E}(x)$ goes to zero in the thermodynamic limit. %Finally, we conjectured that in the quantum ergodic case time integral of the expectation value variance \eqref{newt} is uniformly bounded by Thouless time.
 
Our results open several new research directions. A uniform bound that constraints the expectation value dynamics for all initial states is a potent result, and it would be important to understand if a similar bound applies more generally, beyond ergodic and non-interacting integrable systems.  Another important question is to establish the bound analytically  by outlining necessary conditions for the inequalities (\ref{inequality1},\ref{inequality2}). It is also likely that a similar bound will apply to other time-averaged quantities,  
\bea
{1\over T} \int_0^T a(t)dt \quad {\rm or} \quad {1\over \pi^{1/2}T} \int_{-\infty}^{\infty} e^{-(t/T)^2} a(t) dt.
\eea
Finally, for  an ergodic many-body system the behavior of function $\Delta {E}(x)$ provides a definition of Thouless time, as the time of macroscopic relaxation dynamics. It would be important to relate this definition to other notions based on the amplitude of matrix elements \cite{d2016quantum,serbyn2017thouless} or spectral properties \cite{chan2018spectral}.

%***connection between macroscopic thermalization time and marix elements; definition of Thouless time, connection with other definitions.
%1/sqrt{t} is not a diffusive tail. 

%appendix exponential behavior of square-wave
%remark about quantum scars and largest value of variance-squared

%\vspace{
\begin{acknowledgments}
{\it Acknowledgments.}
I am grateful to Joel Lebowitz and David Huse for helpful discussions.
I would also like to thank Tomaz Prosen, Hong Liu, Moshe Rozali, and Luis Pedro Garcia-Pintos for reading the manuscript and helpful comments. 
I acknowledge the University of Kentucky Center for Computational Sciences for computing time on the Lipscomb High Performance Computing Cluster.
\end{acknowledgments}

\section{Appendix}
\subsection{States that deviate from equilibrium at $t=0$ and any given moment of time $t=t^*$}

We would like to show that there are always special states $\Psi$ which are out-of-equilibrium at the initial moment of time $t=0$ {\it and} at any given moment ${t^*}$. We first show that heuristically and then give a rigorous argument in the Theorem 1 below. 

Let us start with an out-of-equilibrium state $\Psi_0$  such that $a_0=\langle \Psi_0|\hat{A}|\Psi_0\rangle\neq 0$. We consider
\bea
\Psi={\Psi_0(0)+\Psi_0(-{t^*})\over \sqrt{2}},
\label{Psi2}
\eea 
where $\Psi_0(t)$ stands for the time-evolved state in the Schroedinger picture. Then \eqref{Psi2} 
is out-of-equilibrium both as $t=0$ and $t={t^*}$, with $a(0)=a_0/2$ and $a({t^*})=a_0/2$. Here we assumed that ${t^*}$ is large enough such that $\Psi_0$ and $\Psi_0({t^*})$ are sufficiently dephased and $\langle \Psi_0|\Psi_0({t^*})\rangle\approx \langle \Psi_0|\hat{A}|\Psi_0({t^*})\rangle\approx0$. The following theorem makes these additional assumptions unnecessary.

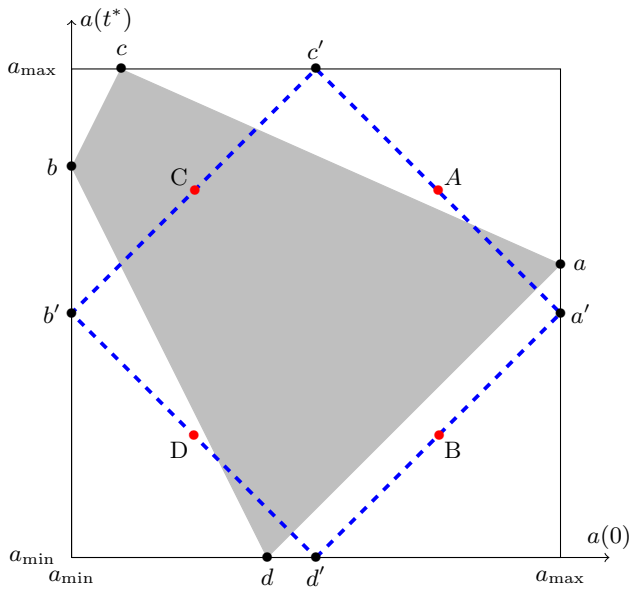
\begin{figure}
\begin{tikzpicture}[scale=1.3]
 %   \coordinate (y) at (1,5);
\draw[->] (1,1) -- (1,6.5) node(yline)[right] {$a({t^*})$};
\draw[->] (1,1) -- (6.5,1) node(xline)[above] {$a(0)$};

 \node [] at (6,.8)  {$a_{\rm max}$};
 \node [] at (1,.8)  {$a_{\rm min}$};
\node [] at (.6,6)  {$a_{\rm max}$};
 \node [] at (.6,1)  {$a_{\rm min}$};

    \draw[] (1,1) -- (1,6) --  (6,6) -- (6,1) -- (1,1);
    \draw[lightgray,fill] (1,5)  -- (3,1) -- (6,4) --(1.5,6);
    \node [black] at (1,5)  {\textbullet};
    \node [black] at (3,1)  {\textbullet};
    \node [black] at  (6,4) {\textbullet};
    \node [black] at (1.51,6) {\textbullet};
    \node[] at(6.2,4) {$a$};
    \node[] at(.8,5) {$b$};
    \node[] at(1.51,6.2) {$c$};
    \node[] at(3,.8) {$d$};

%%%%%%%%%%
  \draw[line width=0.5mm,dashed,blue] (3.5,1) -- (6,3.5) -- (3.5,6)--(1,3.5)--(3.5,1);
    \node [black] at (3.5,1) (a) {\textbullet};
    \node [black] at (3.5,6) (b) {\textbullet};
    \node [black] at (1,3.5) (c) {\textbullet};
    \node [black] at (6,3.5) (d) {\textbullet};
    \node[] at(6.21,3.5) {$a'$};
    \node[] at(.8,3.5) {$b'$};
    \node[] at(3.5,.81) {$d'$};
    \node[] at(3.5,6.2) {$c'$};

    \node [red] at (2.25,2.25) (A) {\textbullet};
    \node [red] at (4.76,2.25) (B) {\textbullet};
    \node [red] at (2.26,4.75) (C) {\textbullet};
    \node [red] at (4.75,4.75) (D) {\textbullet};
    \node[] at(4.9,4.9) {$A$};
    \node[] at(4.9,2.1) {B};
    \node[] at(2.1,4.9) {C};
    \node[] at(2.1,2.1) {D};
\end{tikzpicture}
\caption{Points $a,b,c,d$ which saturate inequalities $a_{\rm max}\ge a(0)\ge a_{\rm min}$ and $a_{\rm max}\ge a({t^*})\ge a_{\rm min}$. Joint numerical range $F$ is convex and therefore includes gray polygon area formed by these four points. Consequently $F$ includes at least one of the four points $A,B,C,D$ highlighted in red. Indeed the only polygon region which does not include $A,B,C,D$ as internal points (although it includes all of them as the boundary points) and touches the boundaries $a(0)=a_{\rm max},\ a(0)=a_{\rm min},\ a({t^*})=a_{\rm max},\ a({t^*})= a_{\rm min}$ is the  blue dotted square formed by the points $a',b',c',d'$. Any attempt to move points $a',b',c',d'$ along the boundaries will make at least one of the points $A,B,C,D$ to be inside $F$.}
\label{fig}
\end{figure}

{\it Theorem 1.} Let $a_{\rm max}$ and $a_{\rm min}$ be the largest and smallest eigenvalues of matrix $\hat{A}$. Then in full generality for any given time ${t^*}$ there is an initial state $\Psi$ such that 
\bea
a(0) \ge {3 a_{\rm max}+a_{\rm min}\over 4} \quad {\rm or} \quad  a(0) \le  {3 a_{\rm min}+a_{\rm max}\over 4},
\eea
and 
\bea
a({t^*}) \ge {3 a_{\rm max}+a_{\rm min}\over 4} \quad {\rm or} \quad  a({t^*}) \le  {3 a_{\rm min}+a_{\rm max}\over 4}.
\eea
{\it Proof.} We would like to to consider a set of points 
\bea
F=\{ (a(0),a({t^*}))\, |\, |\Psi|=1\}.
\eea
 defined for all normalized vectors $\Psi$. $F$ is a compact set which is fully contained within a square region specified by inequalities $a_{\rm max}\ge a(0)\ge a_{\rm min}$ and $a_{\rm max}\ge a({t^*})\ge a_{\rm min}$. There are points in $F$ which saturate each of these four inequalities (we denote them $a,b,c,d$ in Fig.~\ref{fig}).
By definition, the set $F$ is a joint numerical range of matrices $\hat{A}(0)$ and $\hat{A}({t^*})$ (in Heisenberg picture). The Toeplitz-Hausdorff theorem  \cite{toeplitz1918algebraische,hausdorff1919wertvorrat,gustafson1970toeplitz} assures that $F$ is convex. From here it follows that one of the points 
\bea
A={(3 a_{\rm max}+a_{\rm min},3 a_{\rm max}+a_{\rm min})\over 4}, \\
B={(3 a_{\rm max}+a_{\rm min},3 a_{\rm min}+a_{\rm max})\over 4}, \\
C={(3 a_{\rm min}+a_{\rm max},3 a_{\rm max}+a_{\rm min})\over 4}, \\
D={(3 a_{\rm min}+a_{\rm max},3 a_{\rm min}+a_{\rm max})\over 4}, 
\eea
belongs to $F$. This finishes the proof. 
As a comment, we notice that this theorem can be formulated for the states of the form \eqref{micro} with any $E, \Delta E$. Then $a_{\rm max}$ and $a_{\rm min}$ would be the maximal and minimal eigenvalues of $\hat{A}$ projected on the corresponding subspace of the full Hilbert space. Since any projection of  $\hat{A}$ is traceless, $a_{\rm max}>0$ and $a_{\rm min}<0$ for all $E,\Delta E$.

The meaning of this theorem is simple: there is always an initial state $\Psi$ which is out of equilibrium at the initial moment of time $t=0$ and at any given moment of time $t={t^*}$. The universal nature of this result  applicable for all quantum systems does not necessarily allow the deviation at the given moment of time $a({t^*})$ to be of the same sign as $a(0)$. Indeed, one can consider a harmonic oscillator with the period $T=2\pi$. Then, independently of the initial conditions $a(\pi)=-a(0)$ where $a(t)$ is the coordinate of the oscillator. 

%Theorem 1 emphasizes the difficulty of defining equilibration time quantum-mechanically. To ignore possible short-lived ``spikes'' of $a(t)$ at late times and assuming the initial state is out-of-equilibrium $a(0)\neq 0$, we can define equilibration time $\tau$ as the maximal time $a(t)$ remains appreciably large (say 50\% of the initial value) after $t=0$,
%\bea
%|a(t)|\ge a(0)/2\quad {\rm for}\quad \tau\ge t\ge 0.
%\eea
%Alternatively we can define $\tau$ as the first instance $t=\tau$ when $a(t)$ becomes exponentially small. To formalize this line of thinking we introduce a third definitions which also takes
%into account that $a(t)$ can grow for some time after $t=0$,
%\bea
%\label{deft}
%\tau:={\int_0^\infty |a(t)| dt\over \max_{t>0} |a(t)|}.
%\eea
%We should note that for e.g.~a weakly damped harmonic oscillator first two definitions would yield $\tau$ approximately equal to the period of oscillations rather than the exponential decay constant. In other words some of our definitions can err on the side or decreasing $\tau$, but it is not crucial for our consideration.  

\subsection{Relation between $\Delta E(x)$, $\Delta \hat{E}(x)$, and connection to canonical universality}
\begin{figure}
\includegraphics[width=0.4\textwidth]{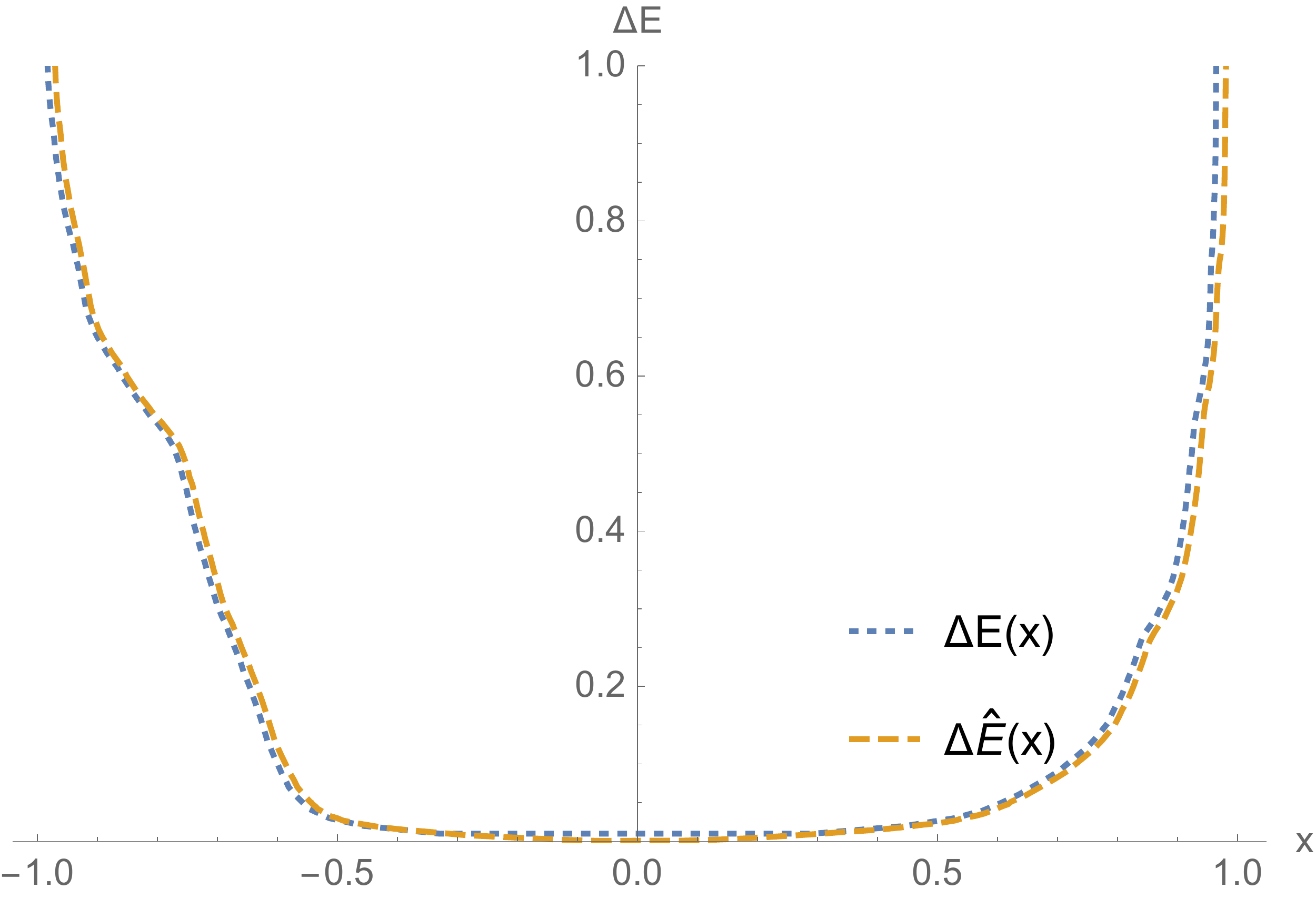}
\caption{Plots of $\Delta {E}(E,x)$  and $\Delta \hat{E}(E,x)$ for $A=\sigma_x^1$, $E=0$ and $h=0.7, L=17$.} 
\label{figcomparison}
\end{figure}

In \cite{dymarsky2017canonical} we have introduced $x_{\rm max/min}(E,\Delta E)$ as the maximal/minimal eigenvalue of an observable $A$ projected on an energy interval $[E-\Delta E,E+\Delta E]$, after subtracting microcanonical expectation value of $A$,
\bea
\label{CU}
x_{\rm max/min}&=&\lambda_{\rm max/min}(A_{ij})-\langle A\rangle_E,\\
\langle A\rangle&=&{1\over {\mathcal N}}\sum_i A_{ii}.
\eea
Here by ${\mathcal N}(\Delta E)$ the total number of energy states inside the interval $[E-\Delta E,E+\Delta E]$, and $\lambda_{\rm max/min}(X)$ stands for the largest/smallest eigenvalue of a Hermitian matrix $X$.
In this paper we introduced $\hat{x}_{\rm max/min}(E,\Delta E)$ of an observable $A$ projected on an energy interval $[E-\Delta E,E+\Delta E]$, after taking all diagonal matrix elements to be zero,
\bea
\hat{x}_{\rm max/min}&=&\lambda_{\rm max/min}(\hat{A}_{ij}).
\eea
Clearly both definitions are closely related. Thus, in the quantum erogidic case when the obserable $A_{ij}$ satisfied the Eigenstate Thermalization Hypothesis ansatz both functions coincide up to unimportant volume-suppressed corrections. Indeed, matrix $A_{ij}$ can be represented as a sum $A_{ij}=\hat{A}_{ij}+D_{ij}$, where $D_{ij}=\delta_{ij}A_{ii}$. Using the triangle inequality we can find
\bea
|x-\hat{x}|\leq ||D_{ij}-\langle A \rangle_E \delta_{ij}||.
\eea
The matrix $D_{ij}-\langle A \rangle_E \delta_{ij}$ is diagonal and its norm is equal to the value of the largest diagonal element, which can be easily estimated to be $\Delta E/{\rm Volume}$ small. Comparing with the behavior $x(\Delta E)\ge f(0)\Delta E^{1/2}$ (see section \ref{smallx} below), we conclude that the contribution of the diagonal elements in the ergodic case is negligible. 
Instead of dealing with the functions $\hat{x}_{\rm max}$ and $\hat{x}_{\rm min}$ and similarly $x_{\rm max/min}$ it is convenient to introduce single-valued inverse functions $\Delta {E}(E,x)$  and $\Delta \hat{E}(E,x)$. 
In the ergodic case both functions are smooth and become zero only at $x=0$, a property dubbed canonical universality in \cite{dymarsky2017canonical}. This is illustrated in Fig.~\ref{figcomparison}.

In the integrable case the behavior of $\Delta {E}(x)$  and $\Delta \hat{E}(x)$ are different as  the contribution of the diagonal matrix elements is not negligible. Both functions are only piece-wise continuous, see Figure~\ref{fig:E0} and Figure~\ref{Z} below.

\subsection{Behavior of $\Delta E(x)$ at small $x$}
\label{smallx}
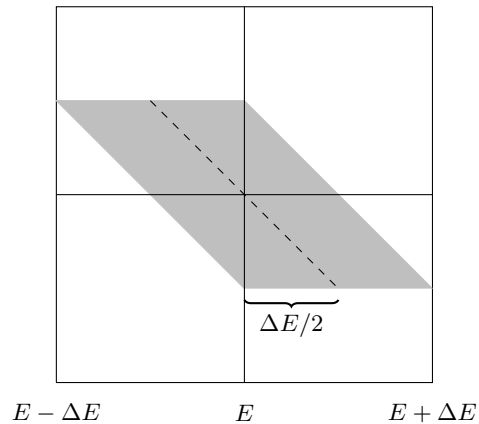
\begin{figure}
\begin{tikzpicture}[scale=1]
 \draw[] (1,1) -- (1,6) --  (6,6) -- (6,1) -- (1,1);
 \draw[lightgray,fill] (3.5,2.25)  -- (1,4.75) -- (3.5,4.75) --(6,2.25);
 \draw[] (3.5,1) -- (3.5,6);
 \draw[] (1,3.5) -- (6,3.5);
 \node[] at(3.5,0.6) {$E$};
 \node[] at(6,0.6) {$E+\Delta E$};
 \node[] at(1,0.6) {$E-\Delta E$};

 \draw[dashed] (2.25,4.75) -- (4.75,2.25);

\draw [
    thick,
    decoration={
        brace,
        mirror,
        raise=0.5cm
    },
    decorate
] (3.5,2.6) -- (4.75,2.6) 
node [pos=0.5,anchor=north,yshift=-0.55cm] {$\Delta E/2$};

\end{tikzpicture}
\caption{Schematic visualization of matrix $\hat{A}_{ij}$ projected inside the interval $E-\Delta E, E+\Delta E$. The integral of the autocorrelation function in \eqref{integral} calculates the sum $\sum_{i,j} |\hat{A}_{ij}|^2$ only over the matrix elements inside the gray area.}
\label{fig2}
\end{figure}

\begin{figure}
\includegraphics[width=0.2\textwidth]{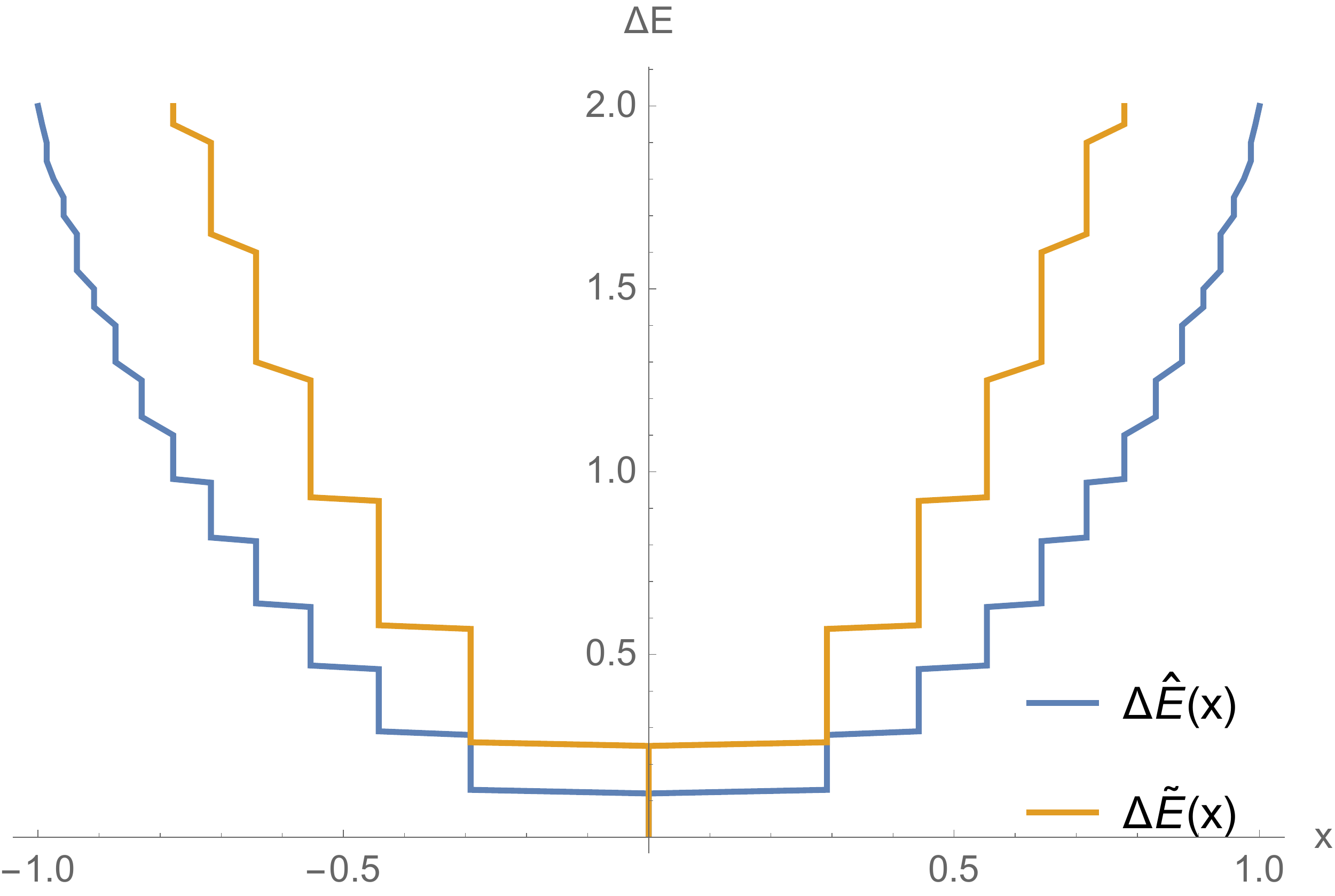}\ 
\includegraphics[width=0.2\textwidth]{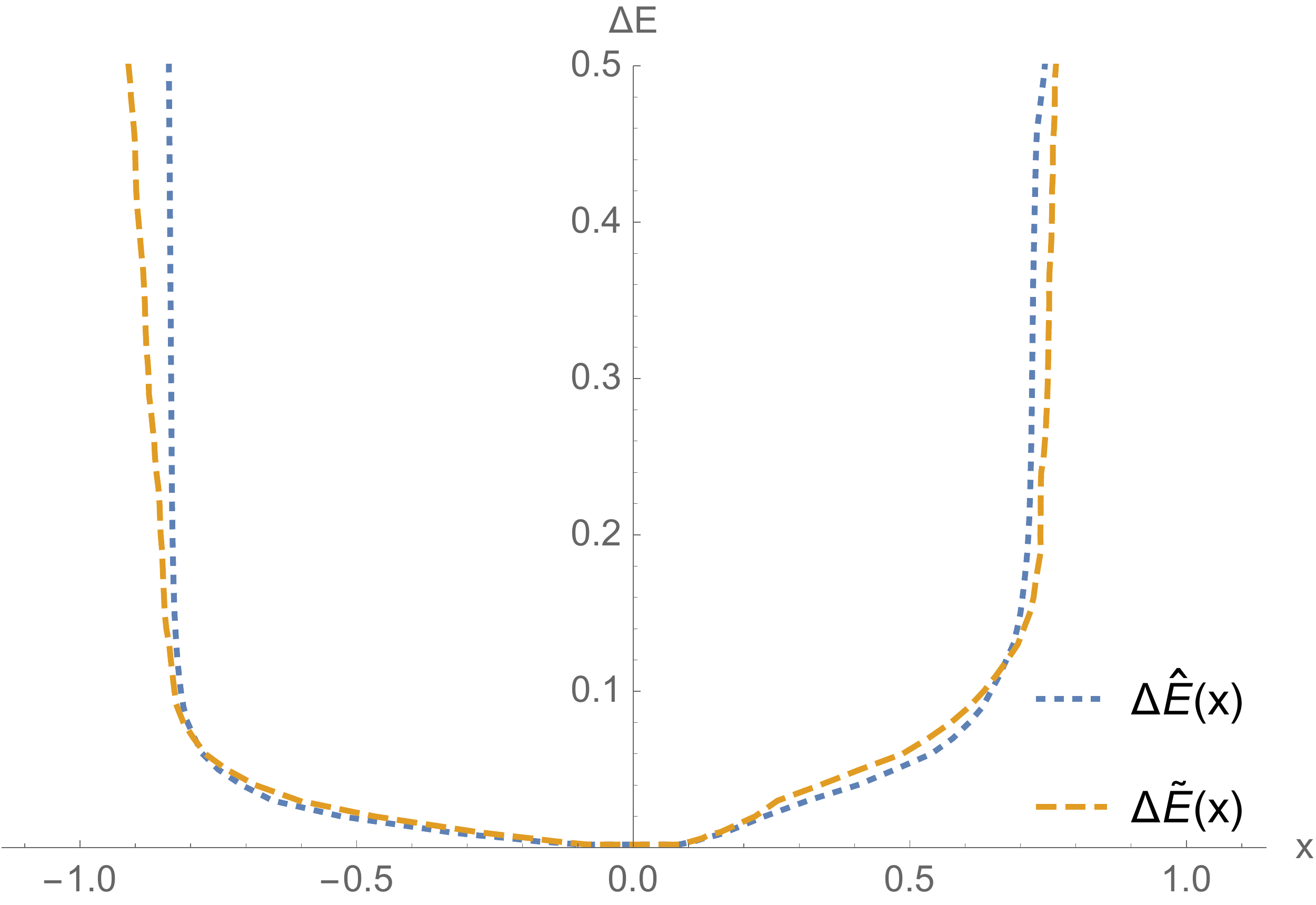}
\caption{$\Delta \hat{E}(x)$ and $\Delta \tilde{E}(x)$ for $A=\sigma_z^1, E=0, L=17$ in the integrable $h=0$ (left) and ergodic $h=0.7$ (right) cases.}
\label{Z}
\end{figure}

\begin{figure}
\includegraphics[width=0.2\textwidth]{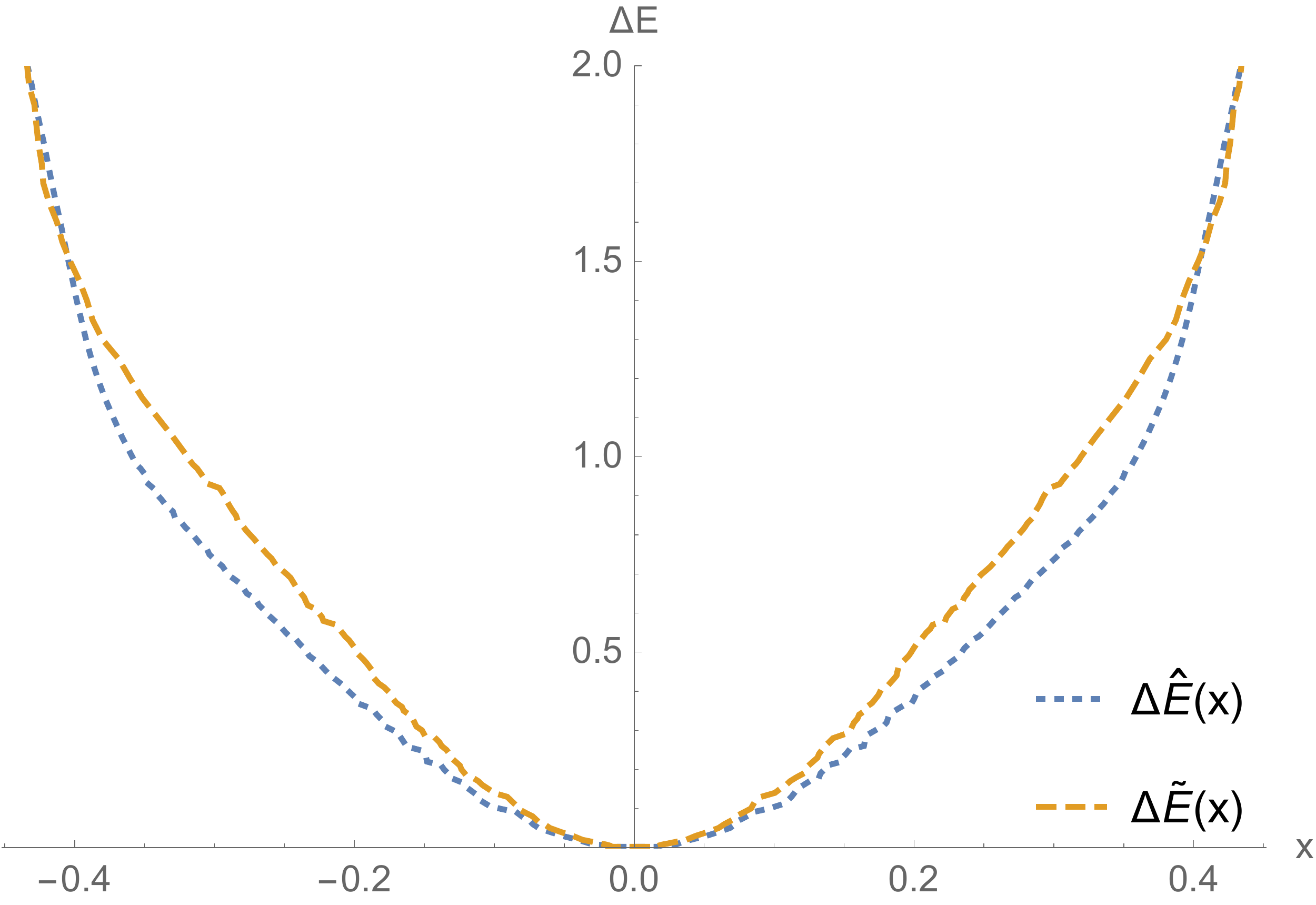}\ 
\includegraphics[width=0.2\textwidth]{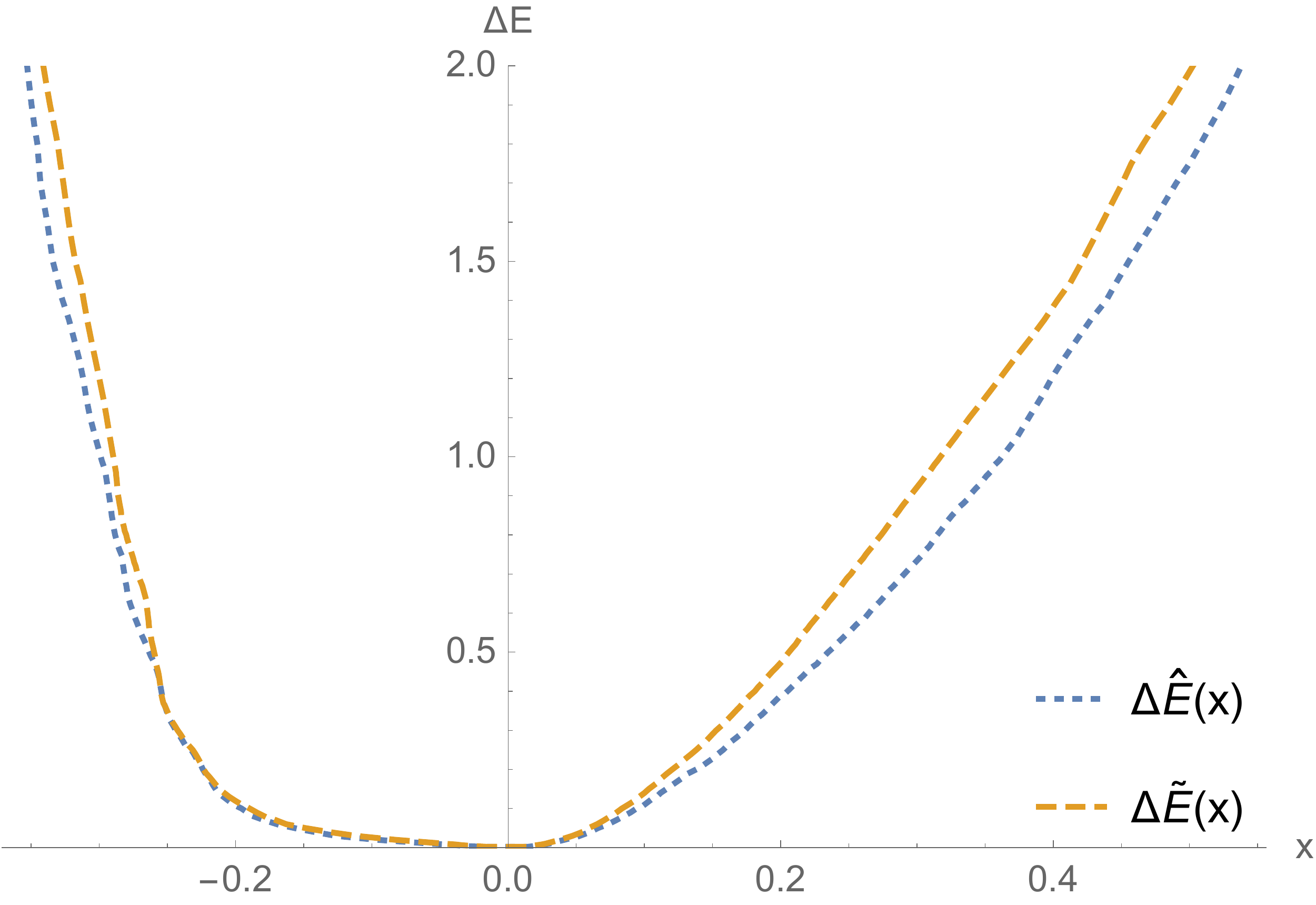}
\caption{$\Delta \hat{E}(x)$ and $\Delta \tilde{E}(x)$ for $A=\sum_i \sigma_z^i/L, E=0, L=17$ in the integrable $h=0$ (left) and ergodic $h=0.7$ (right) cases.}
\label{SumZ}
\end{figure}

For any ${\mathcal N}\times {\mathcal N}$ Hermitian matrix $A_{ij}$ maximal (by absolute value) eigenvalue $\hat{x}$ is bounded by ${\mathcal N}\hat{x}^2 \ge {\rm Tr}{A}{A}^\dagger =\sum_{i,j} |{A}_{ij}|^2$. Let us consider a projection of the full matrix $\hat{A}_{ij}$ of an observable $\hat{A}$ on the energy interval $[E-\Delta E, E+\Delta E]$. By ${\mathcal N}(\Delta E)$ we denote the total number of energy states inside the interval. Then 
\bea
\hat{x}(\Delta E)^2\ge {1\over {\mathcal N}(\Delta E)}\sum_{i,j} |\hat{A}_{ij}|^2,\ E_i,E_j\in [E-\Delta E, E+\Delta E].\nonumber\\
\label{inequality}
\eea 
Let us introduce an autocorrelator of $A$ defined as an average over the microcanonical window $[E-\Delta E, E+\Delta E]$,
\bea
\langle A(t)A(0)\rangle_{\Delta E}={1\over {\mathcal N}}\sum_i \left(\langle E_i|A(t) A(0)|E_i\rangle-|\langle E_i|A|E_i\rangle|^2\right).
\nonumber
\eea
The second term in the parenthesis ensures that $\langle A(t)A(0)\rangle_{\Delta E}$ approaches zero at late $t$.
Then 
\bea
\hat{x}(\Delta E)^2\ge  {{\mathcal N}(\Delta E/2)\over \pi {\mathcal N}(\Delta E)}\int_{-\infty}^{\infty} dt {\sin(t\Delta E /2)\over t} \langle A(t)A(0)\rangle_{\Delta E/2},
\nonumber\\
\label{integral}
\eea
where we substituted the sum in \eqref{inequality} by a partial sum over the matrix elements shown in the gray area in Fig.~\ref{fig2}.
Assuming $\Delta E$ is substantially small such that the density of states $\Omega(E)$ is almost constant within the interval, ${\mathcal N}(\Delta E/2)/ {\mathcal N}(\Delta E) \sim 1/2$, and we find 
\bea
\hat{x}(\Delta E)^2\ge  {1\over 2\pi}\int_{-\infty}^{\infty} dt {\sin(t\Delta E /2)\over t} \langle A(t)A(0)\rangle_{\Delta E/2}.\ \
\label{finalinequality}
\eea
For a many-body quantum system the autocorrelation function $\langle A(t)A(0)\rangle_{\Delta E/2}$ will not depend on the size of the microcanonical window provided 
it includes exponentially many states, $\Delta E\gg \Omega^{-1}$. This puts the bound on how small $\Delta E$ we can consider. When $\Delta E$ is much smaller than the inverse timescale associated with saturation of $\langle A(t)A(0)\rangle$ (normally it would remain finite or grow polynomially with the system size), the integral in \eqref{finalinequality} can be approximated as 
\bea
\label{inequalityhat}
\hat{x}^2\ge  {\Delta Ef^2(0)\over 4\pi}\ ,\\
f^2(0):= \int_{-\infty}^\infty dt  \langle A(t)A(0)\rangle_{\Delta E/2}.\label{fdef}
\eea
The latter integral approaches a constant or grows polynomially with the system size for most observables. Here we define $f^2(0)$ via \eqref{fdef}, but for the quantum ergodic systems satisfying the ETH \cite{d2016quantum} function $f(\omega)$ has an independent meaning. In the latter case the bound \eqref{finalinequality} can be somewhat improved by expressing the full sum from \eqref{inequality} in terms of $f^2(\omega)$. We refer the reader to \cite{dymarsky2017canonical} for details, including  the numerical plots of $\sum_{i,j} |\hat{A}_{ij}|^2/{\mathcal N^2}$ as a function of $\Delta E$ in case of a non-integrable spin chain.

The inequality \eqref{inequalityhat} can be conveniently rewritten as 
\bea
\Delta \hat{E}(x)\leq {4\pi x^2\over f^2(0)},
\eea
but the same inequality also applies to $\Delta E(x)$ \eqref{CU} and $\Delta \tilde{E}(x)$. This is because in the former case the sum in \eqref{inequality} would also include the contribution of the diagonal matrix elements, which is non-negative. In the latter case the sum in \eqref{inequality} should only include matrix elements satisfying $|E_i-E_j|\leq \Delta E$. This condition is actually satisfied by all matrix elements implicitly summed up in \eqref{integral}, which is evident from Fig.~\ref{fig2}.

\begin{figure}
\includegraphics[width=0.2\textwidth]{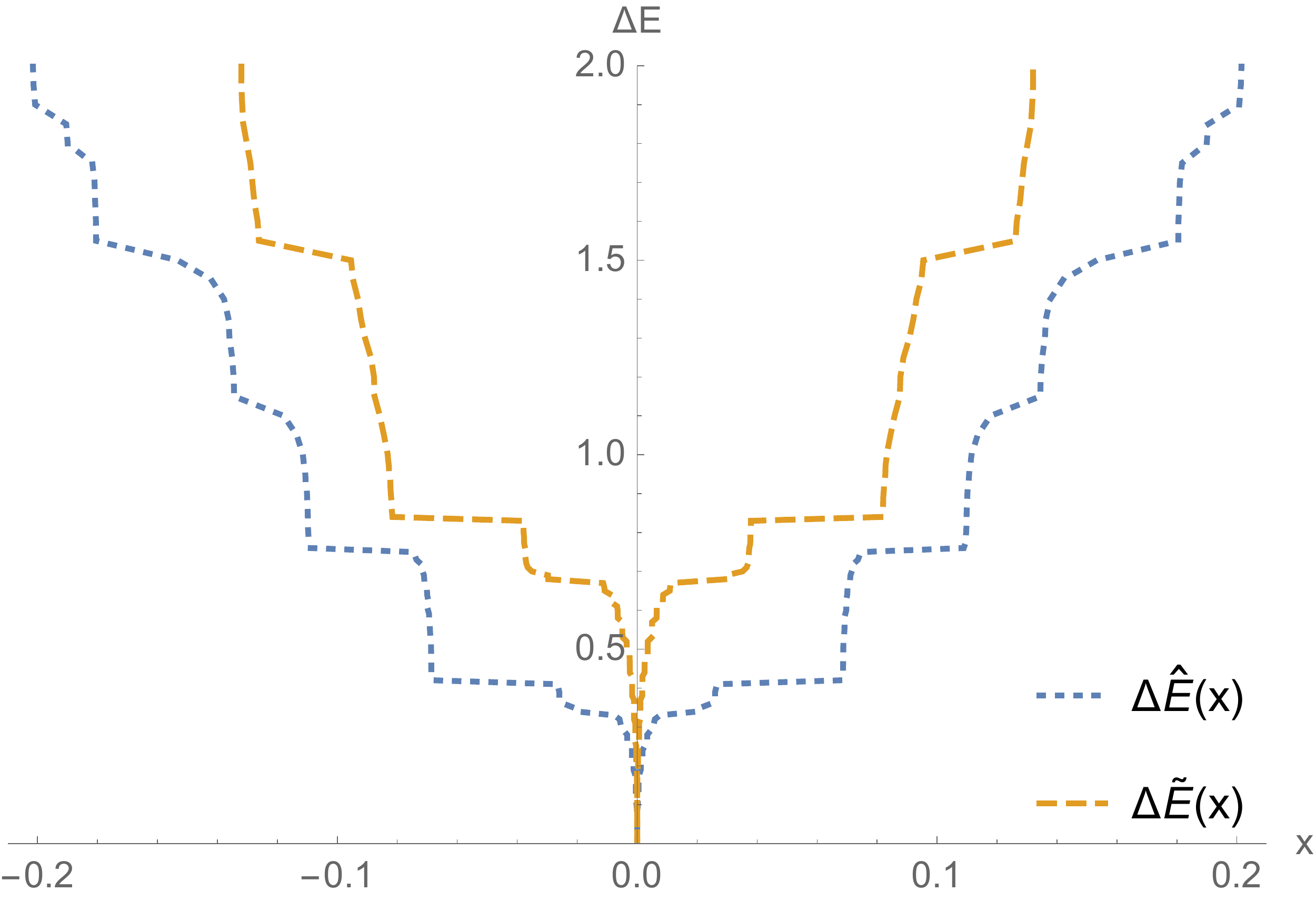}\ 
\includegraphics[width=0.2\textwidth]{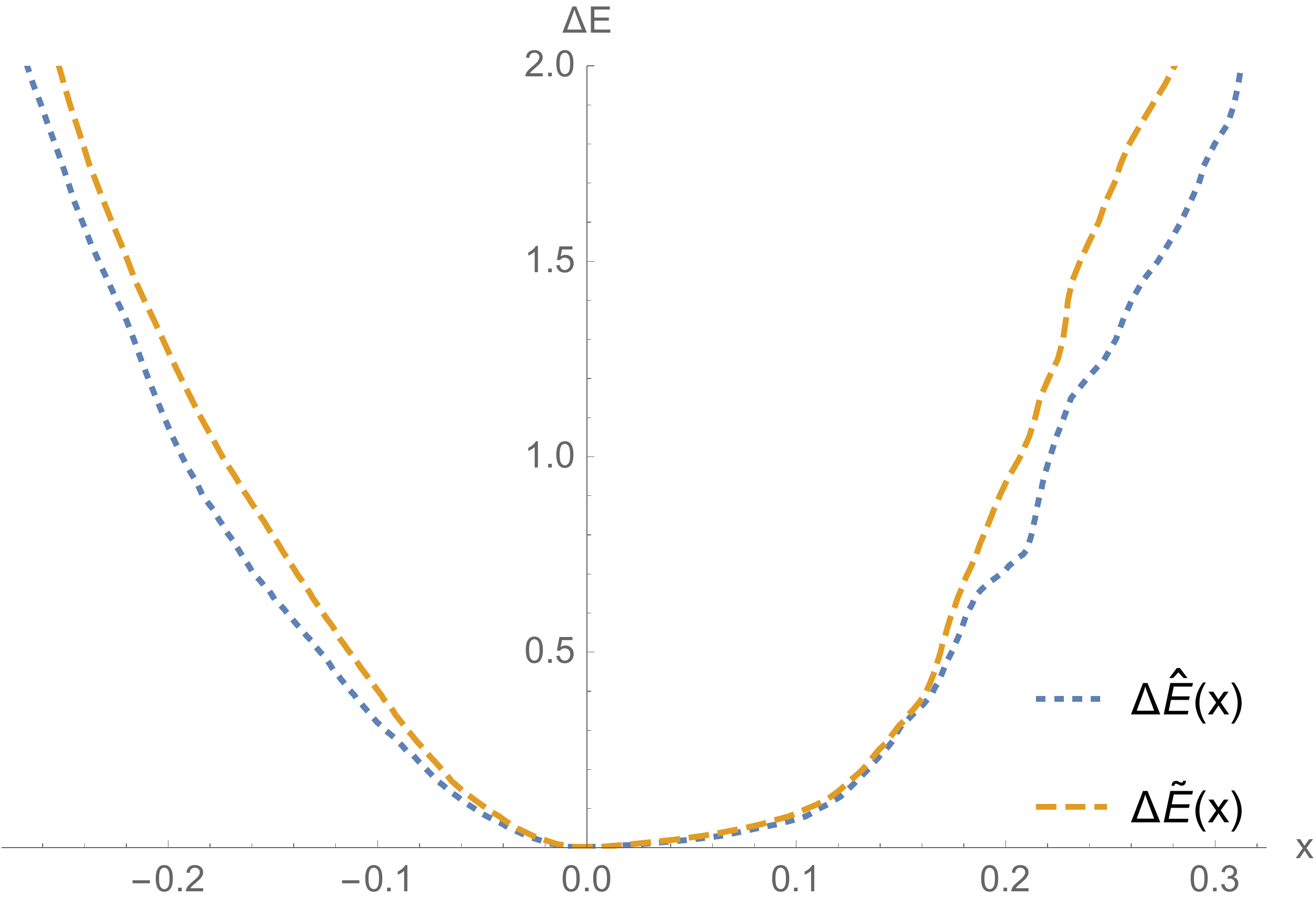}
\caption{$\Delta \hat{E}(x)$ and $\Delta \tilde{E}(x)$ for $A=\sum_i \sigma_x^i/L, E=0, L=17$ in the integrable $h=0$ (left) and ergodic $h=0.7$ (right) cases.}
\label{SumX}
\end{figure}

\subsection{Other operators}
We start with a local operator $A=\sigma_z^1$. The plots or $\Delta \hat{E}(x)$, $\Delta \tilde{E}(x)$ for integrable $h=0$ and ergodic $h=0.7$ cases are shown in Fig.~\ref{Z}.  In the ergodic case the inequality $\Delta \hat{E}(x)\leq \Delta \tilde{E}(x)$  is satisfied for $\Delta E\lesssim \pi/\tau\approx 0.04$, hence \eqref{bound} is valid for $T\geq \tau$, similarly to the case of $A=\sigma_x^1$.

The plots for averaged operators are shown in Figure~\ref{SumZ} and Figure~\ref{SumX}.

\bibliography{eth,equilibrtiondynamics}
%\bibliography{eth,equilibrtiondynamics}
\end{document}